\documentclass[journal]{IEEEtran}

\usepackage[utf8]{inputenc}
\usepackage{graphics,graphicx,color,epsfig}
\usepackage{amsmath,amssymb,amsthm,amsfonts,bm,bbm}
\usepackage{algorithmicx,algorithm,algpseudocode}
\usepackage{multirow}
\usepackage{cite}
\usepackage[normalem]{ulem}

\usepackage{tikz}
\usetikzlibrary{automata, positioning, arrows,shapes.multipart,fit}
\usepackage{pgfplots}
\usepackage{pgfplotstable}
\usepackage{color,amsmath,amssymb,amsthm,epsfig,mathrsfs,cite,bm}
\usepackage{algpseudocode}
\usepackage[utf8]{inputenc}
\usepackage{amsmath}
\usepackage{amsfonts}
\usepackage{placeins}
\usepackage{caption}
\usepackage{bbm,bm}
\usepackage{graphicx,graphics,subcaption}
\usepackage{tikz}
\usepackage{multirow}
\usepackage{pgfplots}
\pgfplotsset{compat=1.15}
\usepgfplotslibrary{fillbetween}
\usetikzlibrary{automata, positioning, arrows,shapes.multipart,fit}
\usepackage[normalem]{ulem}
\usepackage[utf8]{inputenc} % allow utf-8 input
\usepackage[T1]{fontenc}    % use 8-bit T1 fonts
\usepackage{url}            % simple URL typesetting
\usepackage{booktabs}       % professional-quality tables
\usepackage{amsfonts}       % blackboard math symbols
\usepackage{nicefrac}       % compact symbols for 1/2, etc.

\DeclareMathOperator*{\argmax}{argmax} 
\DeclareMathOperator*{\argmin}{argmin}

\newcommand{\revDel}[1]{}

\pgfplotsset{ignore legend/.style={every axis legend/.code={\renewcommand\addlegendentry[2][]{}}}}

\title{BayesAoA: A Bayesian method for Computation Efficient Angle of Arrival Estimation}
\author{%
Akshay Sharma$^*$\thanks{$^*$ equal contribution has been made by the authors} \hspace{8pt} Nancy Nayak$^*$ \hspace{8pt} Sheetal Kalyani\\
\thanks{The authors are with the Department of Electrical Engineering, Indian Institute of Technology Madras, India. \\
Emails: \{ee18s008@smail,ee17d408@smail,skalyani@ee\}.iitm.ac.in}
}
\begin{document}
	\maketitle
	\begin{abstract}
		The angle of Arrival (AoA) estimation is of great interest in modern communication systems. Traditional maximum likelihood-based iterative algorithms are sensitive to initialization and cannot be used online. We propose a Bayesian method to find AoA that is insensitive towards initialization. The proposed method is less complex and needs fewer computing resources than traditional deep learning-based methods. It has a faster convergence than the brute-force methods. Further, a Hedge type solution is proposed that helps to deploy the method online to handle the situations where the channel noise and antenna configuration in the receiver change over time. The proposed method achieves $92\%$ accuracy in a channel of noise variance $10^{-6}$ with $19.3\%$ of the brute-force method's computation.
	\end{abstract}
	
	
	\section{Introduction}
The angle of Arrival (AoA) estimation is one of the location positioning techniques that is used not only in civilian and military domains but also in wideband satellite and wireless cellular communication etc. Real-time accurate AoA estimation helps to reduce the complexity of beamforming for massive multiple-input multiple-output (MIMO) systems. The simplest AoA technique is conventional beamformer (CBF) \cite{van2004optimum}. To improve it, a minimum variance distortionless response (MVDR) beamformer was proposed \cite{van2004optimum,capon1969high}. However, both CBF and MVDR show poor performance in the multi-path environment. The angular resolution is further improved by using the methods like MUltiple Signal Identification and Classification (MUSIC)\cite{schmidt1986multiple}, Root-MUSIC\cite{barabell1983improving}, and Estimation of Signal Parameters via Rotational Invariance Technique (ESPIRIT)\cite{roy1989esprit} that use the signal subspace decomposition (SD) techniques. These methods have poor performance in the presence of multipath and this is overcome by using a smoothing technique that considers the uncorrelated and statistically independent received signals\cite{shan1985spatial}. To handle the correlated received signal in a better way, a parameter estimation algorithm based on maximum likelihood (ML) was proposed in \cite{ziskind1988maximum}. Later two popular ML-based algorithms Expectation Maximization (EM) and Space Alternating Generalized EM (SAGE) were proposed in \cite{chung2002doa}. However, the iterative parametric search takes time to converge. The ML-based parametric estimation methods are highly sensitive to initialization as well. Sparsity-based minimization  \cite{malioutov2005sparse} such as FOCUSS and compressive beamforming for AoA estimation of coherent received signals typically involve a convex optimization problem and hence requires significant computing power. To find the angular response of the received signal, an angular location estimation algorithm based on a probabilistic model was proposed in \cite{bnilam2020angle} and compared with existing AoA techniques. AoA estimation for long-range (LoRa) communication was proposed in \cite{bnilam2020loray}. Recently machine learning methods have been used to estimate AoA and have shown promising results. The learned model of deep learning (DL) based AoA estimation method \cite{huang2018deep} has many parameters thus leading to high complexity. The resource-constrained devices at the receiver side may not have sufficient computing power to deploy DL-based AoA methods. Furthermore, for different noise variances and different antenna configurations at the receiver, the model needs to be retrained and hence cannot be used online.

In this paper, we propose a lightweight Bayesian method to estimate AoA using the Tree Parzen Estimator that is indifferent toward the initial estimate. We then adopt a procedure for early stopping of the proposed method leading to significant savings in the computation. Finally, a Hedge type solution is proposed that does not need a manual tuning of the hyperparameters of the algorithm. It finds the AoAs depending on the channel noise and antenna configuration on the fly thus enabling the algorithm to work online without any retraining. The proposed method achieves $92\%$ accuracy in AoA estimates with a channel noise variance of $10^{-6}$ and an antenna configuration of $8$ in the receiver with just $19.3\%$ of the brute-force method's computation. We also compare our proposed method with the existing MLE-based methods like EM and SAGE. 

	\section{System model}
Consider $M$ far-field signal sources at $\boldsymbol{\theta} = [\theta_1, \dots, \theta_m, \dots, \theta_M]$ and $N$ Uniform Linear Array (ULA) antennas at the receiver. The received signal $\mathbf{z}\in\mathbb{R}^{N\times 1}$ is 
\begin{equation}
\label{eq:sysmodel}
    \mathbf{z} = \mathbf{D}\mathbf{r} + \boldsymbol\nu
\end{equation}
where $\boldsymbol \nu \in \mathbb{R}^{N\times1}$ is the noise vector which is Gaussian with variance $\sigma^2$, $\mathbf{D}\in\mathbb{R}^{N\times M}$ is the steering matrix with each column as $\mathbf{d}(\theta_{m})=\left [{ \begin{matrix} e^{i \frac {2 \pi }{\lambda } d_{1} \sin \left ({\theta _{m}}\right)}, \dots,
    e^{i \frac {2 \pi }{\lambda } d_{N} \sin \left ({\theta _{m}}\right)}  
    \end{matrix} }\right]^T$, $m \in \{1,\dots ,M\}$, and $\mathbf{r}\in\mathbb{R}^{M \times 1}$ is the vector of amplitude values of the received signal corresponding to each source \cite{bnilam2020loray}. Each column $\mathbf{d}(\theta_{m})$ of $\mathbf{D}$ represents the received signal vector at the antenna array with the altered phase corresponding to the $\theta_{m}$ of the $m^{th}$ source. Finding $\mathbf{Dr}$ from the observed data $\mathbf{z}$ can be formulated as an MLE optimization problem as shown in \cite{chung2002doa} and can be solved by either EM or SAGE. For the AoA estimation, the parameters to be estimated are $\boldsymbol{\Gamma} = [\boldsymbol{\Gamma}_1,\dots,\boldsymbol{\Gamma}_m,\dots,\boldsymbol{\Gamma}_M]$ where $\boldsymbol{\Gamma}_m = [\theta_m, r_m]$.
To find the MLE of $\theta$, maximize the likelihood $l_{\mathbf{z}}(\boldsymbol{\theta}) = -\log tr[(\mathbf{I}-\mathbf{P(\theta)})\mathbf{R}_{\mathbf{z}}]
$ where $\mathbf{P(\theta)}=\mathbf{D}(\boldsymbol\theta)(\mathbf{D}^{H}(\boldsymbol\theta)\mathbf{D}(\boldsymbol\theta))^{-1}\mathbf{D}^{H}(\boldsymbol\theta)$ and $\mathbf{R}_{\mathbf{z}}=\mathbf{z}\mathbf{z}^H$ denotes second moment of the samples \cite{Haykin1991AdvancesIS}. Optimization of $l_{\mathbf{z}}(\boldsymbol{\theta})$ involves an $M$ dimensional search over the parameter space $\Theta=\Theta_1\times\Theta_2\times\dots\times\Theta_M$. However, maximizing the likelihood function by doing an $M$-dimensional search is computationally complicated. Therefore, in EM and SAGE, a series of iterations are done to perform single dimensional search instead of a single complex step of finding the MLE.

\begin{algorithm}[t] 
    \caption{EM for AoA estimation}
    \begin{algorithmic}[1]
        \State \textbf{Input:} $\Gamma^0$
        \State $k \leftarrow{}1$
        \While {$\Gamma^k - \Gamma^{k-1} > \epsilon$}
            \State \textbf{E-step:}
                \For {$m\leftarrow{}1$ to $M$}
                    \State Calculate $\mathbf{y}_m$ and $\mathbf{R}_m$
                \EndFor
            \State \textbf{M-step:}
                \For {$m\leftarrow{}1$ to $M$}
                    \State Estimate $\boldsymbol{\Gamma^m}$
                \EndFor
            \State $k\leftarrow{}k+1$
        \EndWhile
    \end{algorithmic}
    \label{alg:EM}
\end{algorithm}
 
\subsection{MLE based methods: EM and SAGE}
For EM, the E-step is,
\begin{equation}
    \begin{aligned}
    \mathbf{z}_m = \mathbf{E}[\mathbf{z}_m|\mathbf{z},\boldsymbol{\Gamma}^t] = \mathbf{d}(\theta_m^t)r_m^t + (\mathbf{z}-\mathbf{D}^t\mathbf{r}^t)
\end{aligned}
\end{equation}
where $\boldsymbol{\Gamma}^t$ is the estimated parameter vector at EM iteration step $t$. The covariance matrix of the $m^{th}$ signal's cross correlation vector is given by $\mathbf{R}_m = \mathbf{z}_m\mathbf{z}_m^H$. The M-step is,
\begin{equation}
    \begin{aligned}
    \theta_m^t &= \argmax_\theta \{\mathbf{d}(\theta)^H\mathbf{R}_m\mathbf{d}(\theta)\} \\
    r_m^t &= \frac{1}{N}  \mathbf{d}(\theta_m^t)^H\mathbf{z}_m.
\end{aligned}
\end{equation}
In case of EM, each EM iteration consists of an E(Expectation)-step and an M(Maximization)-step for all $M$ sources a shown in Alg. \ref{alg:EM}. An alternative of EM called SAGE (Space Alternating Generalized EM) \cite{fessler1994space} was proposed in \cite{chung2002doa} for the purpose of AoA estimation. Instead of carrying out E and M steps for all the sources together, SAGE updates the parameters after the EM step for each source leading to faster convergence. The main challenge of using EM and SAGE is that both these algorithms are very sensitive towards the parameter initialization. As shown in Table \ref{tab:EMSAGE}, a good initialization point produces high accuracy and requires low number of iterations whereas the random initialization shows poor accuracy. Furthermore, a good initial point for a receiver with $N=8$ antennas may not be good for the receiver with $N=6$ antennas as shown in Table \ref{tab:EMSAGE_tab4} where for all the values of $N$, we have used that initialization which is ideal for $N=8$.
\begin{table}[t]
    \centering
    \begin{tabular}{ |c|c|c|c|c|c| } 
    \hline
    Alg. & Initialization & Acc. in $r$ (\%) & Acc. in $\boldsymbol{\theta}$ (\%) & Converges at \\
    \hline
    \multirow{2}{4em}{EM} & Good & $100$ & $100$ & $53$ \\ 
    & Random & $28$ & $18$ & $109$ \\
    \hline
    \multirow{2}{4em}{SAGE} & Good & $100$ & $100$ & $8$ \\
    & Random & $29$ & $31$ & $166$ \\
    \hline
    \end{tabular}
    \caption{Effect of initialization Accuracy with EM and SAGE for $N = 8$, $M = 3$ and $\sigma^2 = 10^{-3}$}
    \label{tab:EMSAGE}
\end{table}

\begin{table}[!t]
    \centering
    \begin{tabular}{ |c|c|c|c|c|c| } 
    \hline
    $N$ & Algorithm & Acc. in $\mathbf{r}$ (\%) & Acc. in $\boldsymbol{\theta}$ (\%) & Converges at \\\hline
    \multirow{2}{1em}{4} & EM & $49$ & $7.9$ & $58$ \\ 
    & SAGE & $8.3$ & $96.5$ & $52$ \\
    \hline
    \multirow{2}{1em}{6} & EM & $98$ & $0$ & $23$ \\ 
    & SAGE & $36$ & $0$ & $476$ \\
    \hline
    \multirow{2}{1em}{8} & EM & $100$ & $100$ & $33$ \\ 
    & SAGE & $100$ & $100$ & $24$ \\
    \hline
    \end{tabular}
    \caption{Accuracy with EM and SAGE for $\sigma^2=10^{-3}$}
    \label{tab:EMSAGE_tab4}
\end{table}

\subsection{Non-linear LS based approach}
Note, the signal model in Eq. \ref{eq:sysmodel} namely, $\mathbf{z} = \mathbf{D}(\boldsymbol{\theta})\mathbf{r} + \boldsymbol\nu$ is linear in $\mathbf{r}$ and non linear in $\boldsymbol{\theta}$. The LS error is $J(\boldsymbol{\theta},\mathbf{r}) = (\mathbf{z}- \mathbf{D}(\boldsymbol{\theta})\mathbf{r})^H(\mathbf{z} - \mathbf{D}(\boldsymbol{\theta})\mathbf{r})$. For a given $\boldsymbol{\theta}$, the $\mathbf{r}$ that minimizes $J(\boldsymbol{\theta},\mathbf{r})$ is given by
\begin{align}
\label{eq:findr}
    \hat{\mathbf{r}} = \left(\mathbf{D}^{H}({\boldsymbol\theta})\mathbf{D}({\boldsymbol\theta})\right)^{-1}\mathbf{D}^{H}({\boldsymbol\theta})\mathbf{z}
\end{align}
The resulting LS error is given by 
\begin{align}
\label{eq:optimizationobj}
    J(\boldsymbol{\theta},\hat{\mathbf{r}}) = \mathbf{z}^H(\mathbf{I}-\mathbf{D}({\boldsymbol\theta})(\mathbf{D}^{H}({\boldsymbol\theta})\mathbf{D}({\boldsymbol\theta}))^{-1}\mathbf{D}^{H}({\boldsymbol\theta}))\mathbf{z}.
\end{align}
Therefore the problem to be solved is 
\begin{equation}
    \begin{aligned}
    \label{eq:optimizer}
    \hat{\boldsymbol{\theta}} &= \argmax_{\theta} f(\boldsymbol\theta) = \mathbf{z}^H\mathbf{P}(\theta)\mathbf{z},
\end{aligned}
\end{equation}
where $\mathbf{P}(\theta)= (\mathbf{D}({\boldsymbol\theta})(\mathbf{D}^{H}({\boldsymbol\theta})\mathbf{D}({\boldsymbol\theta}))^{-1}\mathbf{D}^{H}({\boldsymbol\theta}))$. Once $\boldsymbol{\theta}$ is estimated using eq. \ref{eq:optimizer}, $\mathbf{r}$ is found just by least squares as given in eq. \ref{eq:findr}. 

\subsubsection{\textbf{Proposed BayesAoA}}
Brute force grid search and random search are well known methods for parameter search but not efficient in cases where the evaluation of objective function is expensive since they do not take the previous evaluations into account to draw the next set of parameters. Note that grid search, random search\cite{bergstra2012random} and particle swarm optimization (PSO) \cite{kennedy1995particle} come under the class of non-Bayesian optimization. We propose to use a Bayesian approach such as Sequential Model Based Optimization (SMBO) method to search for the set of angles $\hat{\boldsymbol{\theta}}=\argmin_{\boldsymbol\theta\in\boldsymbol\Theta} f(\boldsymbol\theta)$ taking the past evaluation with old parameters into account. We call the proposed method based on SMBO as \textit{BayesAoA}. The proposed method first initializes a probabilistic regression model $\mathcal{F}$ by randomly sampling a very small set of observations $\{\boldsymbol\theta_1, \dots, \boldsymbol\theta_i\}$. The model $\mathcal{F}$ maps the AoAs to the score/probability of the objective function given by $p(f|\boldsymbol\theta)$ that works as a cheap surrogate of the expensive objective function $ f(\boldsymbol\theta)$. Post initialization, a new set of AoAs $\boldsymbol\theta_{i+1}\in\boldsymbol\Theta$ is found that gives the best selection function $S$. The selection function $S$ used here is Expected Improvement (EI) and is given by 
\begin{align}
    EI_{f^*}(\boldsymbol\theta) = \int_{-\infty}^{f^*} (f^*-f)p(f|\boldsymbol\theta)df
\end{align}
where $f^*$ is some threshold of the objective function, $f$ is the value of true objective function for AoA set $\boldsymbol\theta$. The true objective function is applied to the optimal AoAs to obtain $f(\boldsymbol\theta_{i+1})$. The surrogate model is then updated with these results $\{\boldsymbol\theta_{i+1},f(\boldsymbol\theta_{i+1})\}$. New set of AoAs generated in the above way is used to update the surrogate model until a maximum amount of iteration $K$ is reached. Based on how the surrogate $p(f|\boldsymbol\theta)$ of the objective and the selection function $S$ is chosen to find the next set of AoAs, different variants of SMBO exist. The aim is to maximize this expected improvement over the domain of AoAs $\boldsymbol\Theta$. If $p(f|\boldsymbol\theta)$ is zero everywhere that $f<f^*$, then the EI for the AoA set $\boldsymbol\theta$ is zero, otherwise, the AoA $\boldsymbol\theta$ has an expected improvement over the threshold $f^*$. Whereas the other SMBO algorithms like Gaussian processes or Random forests use a predictive distribution like $p(f|\boldsymbol\theta)$, the Tree Parzen Estimator\cite{bergstra2011algorithms} (TPE) that is used in the proposed BayesAoA algorithm, models $p(\boldsymbol\theta|f)$ and $p(f)$ to obtain $p(f|\boldsymbol\theta)$ and is given by 
\begin{align}
\label{eq:bayespost}
    p(f|\boldsymbol\theta) = \frac{p(\boldsymbol\theta|f)p(f)}{p(\boldsymbol\theta)}
\end{align}
where $p(\boldsymbol\theta)$ is the prior density of the AoAs. For TPE, it can be Gaussians with means centered at $\{\boldsymbol\theta_1,\dots.\boldsymbol\theta_i\}$ and standard deviation set to the greater of distances to the left and right neighbour but clipped to stay in a reasonable range. The probability of the AoAs given the score of the objective function $p(\boldsymbol\theta|f)$ is given by
\begin{align}
\label{eq:tpe}
    p(\boldsymbol\theta|f) = \begin{cases}
                l(\boldsymbol\theta), & \text{if } f<f^*\\
                g(\boldsymbol\theta),  & \text{if } f\geq f^*
                \end{cases}.
\end{align} 
\begin{algorithm}[!t] 
    \caption{BayesAoA}
    \begin{algorithmic}[1]
        \State \textbf{Parameters:} Number of sources $M$, number of receivers $N$, max number of iteration $K$, TPE quantile $\gamma$.
        \State Initialize $\bar{\boldsymbol{\theta}}_{0}=\{\boldsymbol{\theta}_1, \boldsymbol{\theta}_2, \dots, \boldsymbol{\theta}_i\}$ where each element of $\bar{\boldsymbol{\theta}}_{0}$ comes from a uniform random distribution.
        \State Create a list $L=\{(\boldsymbol{\theta}_0, f(\boldsymbol{\theta}_0)), \dots, (\boldsymbol{\theta}_i, f(\boldsymbol{\theta}_i))\} $ 
        \For {$k= i+1,\ldots, K$}
            \State Using $L$ and $\gamma$, find $p(\boldsymbol{\theta}|f)$ by eq. (\ref{eq:tpe}).
            \State Find $p(f|\boldsymbol\theta)$ using eq. (\ref{eq:bayespost}).
            \State $\boldsymbol{\theta}_{k}\leftarrow{} \argmax_{\boldsymbol\theta\in \boldsymbol\Theta} EI_{f^*}(\boldsymbol\theta)$
            \State Append $(\boldsymbol{\theta}_{k}, f(\boldsymbol{\theta}_{k}))$ to $L$.
        \EndFor
        \State Return $\boldsymbol{\theta}_{K}$
    \end{algorithmic}
    \label{alg:BayesAoA}
\end{algorithm}
Here, $l(\boldsymbol\theta)$ is the density formed by the samples $\{\boldsymbol\theta_i\}$ such that the objective functions $\{f(\boldsymbol\theta_i)\}$ are lesser than the threshold $f^*$ and $g(\boldsymbol\theta)$ is formed by the remaining samples. TPE maximizes the expected improvement given by
\begin{align}
    EI_{f^*}(\boldsymbol\theta) &= \frac{\gamma f^* l(\boldsymbol\theta)-l(\boldsymbol\theta)\int_{-\infty}^{f^*}p(f)df}{\gamma l(\boldsymbol\theta)+(1-\gamma)g(\boldsymbol\theta)} \\
    &\propto \left(\gamma+\frac{g(\boldsymbol\theta)}{l(\boldsymbol\theta)}(1-\gamma)\right)^{-1},
\end{align}
where the TPE algorithm chooses threshold $f^*$ to be some quantile $\gamma$ of the observed $f(\boldsymbol\theta)$s so that $p(f<f^*)=\gamma$. This implies that, to improve the expected improvement, the points $\boldsymbol\theta_i$ should have more probability under the density $l(\boldsymbol\theta)$ and less under $g(\boldsymbol\theta)$. Based on domain knowledge, we know that no two distinct sources can have the same AoA and we incorporate that into the algorithm that is detailed in Alg. \ref{alg:BayesAoA}.

\begin{algorithm}[!t] 
    \caption{BayesAoA-ES}
    \begin{algorithmic}[1]
        \State \textbf{Parameters:} Number of source $M$, number of receivers $N$, max number of iteration $K$, TPE quantile $\gamma$, gradient threshold $\epsilon_g$, ES interval: $I$.
         \State Initialize $\bar{\boldsymbol{\theta}}_{0}=\{\boldsymbol{\theta}_1, \boldsymbol{\theta}_2, \dots, \boldsymbol{\theta}_i\}$ where each element of $\bar{\boldsymbol{\theta}}_{0}$ comes from a uniform random distribution.
        \State Create a list $L=\{(\boldsymbol{\theta}_0, f(\boldsymbol{\theta}_0)), \dots, (\boldsymbol{\theta}_i, f(\boldsymbol{\theta}_i))\} $ 
        \For {$k= i+1,\ldots, K$}
            \State Using $L$ and $\gamma$, find $p(\boldsymbol{\theta}|f)$ by eq. (\ref{eq:tpe}).
            \State Find $p(f|\boldsymbol\theta)$ using eq. (\ref{eq:bayespost}).
            \State $\boldsymbol{\theta}_{k}\leftarrow{} \argmax_{\boldsymbol\theta\in \boldsymbol\Theta} EI_{f^*}(\boldsymbol\theta)$
            \If{($k\%I=0$ and max($Grad(f,\boldsymbol{\theta}_{k})$) $\leq \epsilon_g$) or $k=K$}
                \State Return $(\boldsymbol{\theta}_{k},k)$.
                
            \Else
                \State Append $(\boldsymbol{\theta}_{k}, f(\boldsymbol{\theta}_{k}))$ to $L$.
            \EndIf
        \EndFor
    \end{algorithmic}
    \label{alg:BayesAoA-ES}
\end{algorithm}

\subsubsection{\textbf{Proposed BayesAoA-ES}}
The proposed BayesAoA algorithm continues to draw samples till the maximum iteration $K$, even though it attains the best selection function $S=EI_{f^*}(\boldsymbol\theta)$ much before $K$. We propose to stop the algorithm BayesAoA as soon as it achieves the optimal set of AoAs based on \textit{Early Stopping} (ES) to save unnecessary computation involved and therefore is called BayesAoA-ES. In this method, at every iteration, the partial derivatives
\begin{align*}
    Grad(f,\boldsymbol{\theta}_k)=\left\{
    \frac{\partial f(\boldsymbol{\theta}_k)}{\partial\theta_{kx}},
    \frac{\partial f(\boldsymbol{\theta}_k)}{\partial\theta_{ky}},
    \frac{\partial f(\boldsymbol{\theta}_k)}{\partial\theta_{kz}}\right\}
\end{align*}
of the multidimensional function $f$ at $\boldsymbol{\theta}_k$ are compared with a threshold $\epsilon_g$. If all of the partial derivatives in $Grad(f,\boldsymbol{\theta}_k)$ are lesser than $\epsilon_g$, the optimal set of AoAs are considered to be reached. The algorithm is detailed in Alg. \ref{alg:BayesAoA-ES}. As the function is not known, calculating the partial derivatives by differentiation is not possible. A method of numerical gradient calculation is used to calculate $Grad(f,\boldsymbol{\theta}_k)$ and is given in Alg. \ref{alg:grad}. Note that the calculation of $Grad(f,\boldsymbol{\theta}_k)$ needs to evaluate $f(\boldsymbol{\theta}_k)$ for $2M$ times at every iteration. So instead the ES condition can be tested with an interval of $I$ iterations. Under this if the BayesAoA-ES algorithm stops at $\hat{k}^{th}$ iteration using ES, only $2M*(\hat{k}/I)$ expensive calculations need to be done across all iterations.

\begin{algorithm}
\caption{$Grad(f,\boldsymbol{\theta}_k)$}
\begin{algorithmic}[1]
    \State $\delta=\boldsymbol{\theta}_k/10000$
    \For{$i=1,\ldots,M$}
        \If{$\boldsymbol{\theta}_k(i)==0$}
            \State $\delta(i)=10^{-12}$
        \EndIf
        \State $u=\boldsymbol{\theta}_k$
        \State $u(i)=\boldsymbol{\theta}_k(i)+\delta(i)$
        \State $f_1 = f(u)$
        \State $u(i)=\boldsymbol{\theta}_k(i)-\delta(i)$
        \State $f_2 = f(u)$
        \State $g(i)=(f_1-f_2)/(2*\delta(i))$
    \EndFor
    \State Return $\mathbf{g}$
\end{algorithmic}
\label{alg:grad}
\end{algorithm}

A challenge in using the proposed algorithm BayesAoA-ES is that the correctness of the estimated AoA varies highly with the chosen threshold $\epsilon_g$. The loss surface varies with number of receivers and so does the gradients around the optimal point. For $N=\{3,4,5\}$, the minima is not sharp enough to qualify as unambiguous. This motivates us to find an optimum gradient threshold $\epsilon_g$ for each receiver configuration.

\subsubsection{\textbf{HedgeBayesAoA-ES}}
We employ a Hedge \cite{freund1997decision} type solution wherein a total of $B$ potential thresholds ($\epsilon_g$) are treated as expert candidates. Equal weights are assigned to all the experts at time step $t=1$. For HedgeBayesAoA-ES, at each timestep $t$, BayesAoA-ES returns $(\boldsymbol\theta_{kb}^{t},k_{b}^{t})$ for each expert with $\epsilon_g$. Depending on the accuracy of the prediction of $\boldsymbol{\theta}_{kb}^{t}$ and the iteration $k_{b}^{t}$ at which the algorithm converges for time step $t$ and expert $b$, the loss metric is calculated as
\begin{align}
    l_{b}^{t} = (1-\zeta)*err_{b}^{t} + \zeta*\hat{k}_{b}^{t}/K.
\end{align}
The weights on the experts are updated as $w_{b}^{t+1}=w_{b}^{t} \cdot \beta^{l^{t}_{b}}$. The algorithm to train and find the optimum threshold is discussed further in Alg. \ref{alg:HedgeBayesAoA}. Over time, the HedgeBayesAoA-ES will give more weight to expert that performs best in terms of both accuracy and average number of iterations and one can deactivate the other experts if the external factors like $N$ and $\sigma^2$ remain unchanged. Furthermore, when the factors like $N$ and $\sigma^2$ change, HedgeBayesAoA-ES will be able to reset the weights and update the weights according to the new $N$ and $\sigma^2$. Note that, there is always a trade off between accuracy and average number of iterations needed for the algorithm to converge. Depending on the application at hand, the value of $\zeta$ can be chosen. A higher value of $\zeta$ will give more importance to accuracy and less importance to average number of iterations needed.
\begin{algorithm}
    \caption{HedgeBayesAoA-ES}
    \begin{algorithmic}[1]
        \State Let the threshold vector be $\bar{\boldsymbol{\epsilon}}_{0}=\{\boldsymbol{\epsilon}_1, \boldsymbol{\epsilon}_2, \dots, \boldsymbol{\epsilon}_B\}$ and its corresponding weights vector be $\boldsymbol{w^{t}} = \{w_{1}^{t}, w_{2}^{t} \dots, w_{B}^{t}\}$. 
        \State Let the actual angles of $M$ sources are $\boldsymbol\theta=\{\theta_1,\dots,\theta_m,\dots,\theta_M\}$, Hedge hyperparameter $\beta$, importance parameter $\zeta$.
        \For {t=$1,2 \ldots, T$}
            \For{b=$1,2 \ldots, B$}
                \State Carry out BayesAoA-ES for $\epsilon_{b}$ and return $(\boldsymbol{\theta}_{kb}^{t},k_{b}^{t})$. 
                \If {$\boldsymbol{\theta}_{kb}^{t} == \boldsymbol\theta$ }
                    \State $err_{b}^{t} = 0$ 
                \Else
                    \State $err_{b}^{t}=1$ 
                \EndIf
                \State  $k_{b}^{t}=k$ 
                \State $l_{b}^{t} = (1-\zeta)*err_{b}^{t} + \zeta*\hat{k}_{b}^{t}/K$.
                \State $w_{b}^{t+1}=w_{b}^{t} \cdot \beta^{l^{t}_{b}}$
            \EndFor 
            \State  $\boldsymbol{p^{t}} = \frac{\boldsymbol{w^{t}}}{\sum^{B}_{b=1}w^{t}_{b}}$ (Normalise $\boldsymbol{w^{t}}$ to obtain $\boldsymbol{p^{t}}$)
        \EndFor
        \end{algorithmic}
        \label{alg:HedgeBayesAoA}
    \end{algorithm}

\begin{table*}[!htb]
\begin{minipage}{.32\linewidth}
\begin{tabular}{| m{1em} | m{0.6cm}| m{1cm} | m{1.3cm} | } 
\hline
$N$ & $\epsilon_{g}$  & Accuracy (\%) & Avg no of iterations \\
\hline
\multirow{3}{2em}{$4$} & $1.0$ & $2$  &$100$  \\
& $0.5$ & $4$  & $200$ \\ 
& $0.1$ & $2$  & $500$ \\
& $0.05$ & $6$  & $600$ \\
& $0.01$ & $8$  & $1000$ \\
\hline
\multirow{3}{2em}{$6$} & $1.0$ & $20$  &$600$  \\
& $0.5$ & $26$  & $600$ \\ 
& $0.1$ & $28$  & $900$ \\
& $0.05$ & $32$  & $1000$ \\
& $0.01$ & $36$  & $1000$ \\
\hline
\multirow{3}{2em}{$8$} & $1.0$ & $38$  &$800$  \\
& $0.5$ & $42$  & $900$ \\ 
& $0.1$ & $56$  & $1000$ \\
& $0.05$ & $54$  & $1000$ \\
& $0.01$ & $64$  & $1000$ \\
\hline
\end{tabular}
\end{minipage}%
\begin{minipage}{.32\linewidth}
\begin{tabular}{ | m{1em} | m{0.6cm}| m{1cm} | m{1.3cm} |  } 
\hline
$N$ & $\epsilon_{g}$  & Accuracy (\%) & Avg no of iterations \\
\hline
\multirow{3}{2em}{$4$} & $1.0$ & $4$  &$100$  \\
& $0.5$ & $16$  & $200$ \\ 
& $0.1$ & $18$  & $400$ \\
& $0.05$ & $18$  & $600$ \\
& $0.01$ & $42$  & $800$ \\
\hline
\multirow{3}{2em}{$6$} & $1.0$ & $42$  &$300$  \\
& $0.5$ & $44$  & $400$ \\ 
& $0.1$ & $50$  & $800$ \\
& $0.05$ & $66$  & $900$ \\
& $0.01$ & $68$  & $900$ \\
\hline
\multirow{3}{2em}{$8$} & $1.0$ & $64$  &$400$  \\
& $0.5$ & $80$  & $700$ \\ 
& $0.1$ & $72$  & $900$ \\
& $0.05$ & $72$  & $1000$ \\
& $0.01$ & $74$  & $1000$ \\
\hline
\end{tabular}
\end{minipage}
\begin{minipage}{.32\linewidth}
\begin{tabular}{ | m{1em} | m{0.6cm}| m{1cm} | m{1.3cm} |  } 
\hline
$N$ & $\epsilon_{g}$  & Accuracy (\%) & Avg no of iterations \\
\hline
\multirow{3}{2em}{$4$} & $1.0$ & $6$  &$100$  \\
& $0.5$ & $8$  & $200$ \\ 
& $0.1$ & $16$  & $400$ \\
& $0.05$ & $20$  & $400$ \\
& $0.01$ & $40$  & $500$ \\
\hline
\multirow{3}{2em}{$6$} & $1.0$ & $42$  &$300$  \\
& $0.5$ & $58$  & $400$ \\ 
& $0.1$ & $60$  & $500$ \\
& $0.05$ & $68$  & $600$ \\
& $0.01$ & $70$  & $900$ \\
\hline
\multirow{3}{2em}{$8$} & $1.0$ & $54$  &$300$  \\
& $0.5$ & $68$  & $400$ \\ 
& $0.1$ & $82$  & $700$ \\
& $0.05$ & $92$  & $900$ \\
& $0.01$ & $94$  & $1000$ \\
\hline
\end{tabular}
\end{minipage}%
\caption{Accuracy and average no of iterations required for BayesAoA-ES algorithm to converge for $M=3$ sources, noise variance $\sigma^2=\{10^{-2}(\text{left}),10^{-4}(\text{middle}),10^{-6}(\text{right})\}$ and number of receivers $N$ and Gradient threshold $\epsilon_g$, $K=1000$.}
\label{tab:BayesAoA-ES}
\end{table*}

\begin{table}[ht]
    \centering
    \begin{tabular}{|c|c|c|}
    \hline
        Method & Accuracy (\%) & Computation  \\
        \hline
        Brute force & $100$ & $4960$ \\
        \hline
        BayesAoA & $90$ & $1000$ \\
        \hline
        BayesAoA-ES($\epsilon_g=0.05$) & $92$ & $900+2*3*9=954$ \\
        \hline
    \end{tabular}
    \caption{Saving in computation by BayesAoA-ES with a decent accuracy with $N=8, \sigma^2=10^{-6}$.}
    \label{tab:computation}
\end{table}

\section{Experimental Setup and results}
The setup for the simulation considers a Uniform Linear Array (ULA) of $4,6$ and $8$ receivers i.e. $(N=4,6,8)$. The number of source transmitters to be detected is set to three $(M=3)$. AWGN is considered with variance ($\sigma^2$) values equal to $10^{-6}$, $10^{-4}$ and $10^{-2}$. The angles of arrival of the source transmitters are chosen from a set of $\theta$s which has a resolution of $0.1$ radian. In other words, any two AoA differ by at least $0.1$ rad. With a resolution of $0.1$ rad, between $-\pi/2$ and $\pi/2$, $\boldsymbol\Theta$ is a set of $32$ values: $\{-1.57, -1.47, \dots, 1.57\}$. Because no two of $3$ sources can take the same value, therefore $\boldsymbol{\theta}$ can have ${32 \choose 3}=4960$ distinct AoA combination. To search over the same set, brute-force method predicts AoAs with $100\%$ accuracy but needs to compute $f(\boldsymbol{\theta})$ for $4960$ times. The two metrics we look at are (i) the accuracy of AoA prediction and (ii) the average number of iterations. The proposed BayesAoA algorithm draws samples such that at every iteration the expected improvement $EI_{f^*}$ improves thus achieving an accuracy of $90\%$ just in $1000$ iterations (in case the algorithm is run for a maximum of $K=1000$ iterations) compared to $4960$ iterations of brute force method. Because, unlike the brute-force method, BayesAoA does not draw samples blindly but by taking the previous evaluations into account. However, it is difficult to set a maximum iteration apriori. So to make the algorithm draw samples only until when the desired accuracy is achieved, BayesAoA-ES is used where the partial derivatives are compared with the threshold at every $I=100^{th}$ iteration.

The proposed method BayesAoA-ES with a carefully chosen $\epsilon_g$ can converge to the accurate $\boldsymbol{\theta}$ much before $4960$ iterations\footnote{here iteration is defined as the number of computation of $f(\boldsymbol{\theta})$ needed for convergence.} as shown in Table \ref{tab:BayesAoA-ES}. Note that, MLE-based EM and SAGE also converge quite fast but only with a good initialization. All the experiments for BayesAoA-ES are given chance to run for a maximum of $K=1000$ iterations usually and the results are averaged over $50$ independent runs. The performance of the proposed methods improve with more number of antennas in the receiver (refer to the performances at $N=\{4,6,8\}$ at $\sigma^2=10^{-6}$) and with decreasing noise variance (refer to the performances at different $\sigma^2$ for $N=6$). Observe that, with a higher value of $\epsilon_g$, the BayesAoA-ES algorithm stops early but achieves a lesser accuracy and vice versa. With a channel noise variance $\sigma^2=10^{-6}$ and for the receiver with $6$ antennas,  the algorithm has an accuracy of $68\%$ and converges at $600$ iterations at $\epsilon_g=0.05$ whereas it has an accuracy of $70\%$ that converges at $900$ iterations. The saving in computation using BayesAoA-ES is $80.7\%$ compared to brute-force with an accuracy of $92\%$ in AoA estimation as shown in Table. \ref{tab:computation}. A proper $\epsilon_g$ should be chosen depending on the requirement of accuracy and the support for computation. In a scenario where the receiver may use a different number of antennas interchangeably, or the noise in the system varies, it is difficult to deploy the AoA algorithms with just a single $\epsilon_g$. To choose an $\epsilon_g$ dynamically based on the current channel condition or the number of receiver antennas, the proposed HedgeBayesAoA-ES algorithm chooses the best $\epsilon_g$. In our experiment, the Hedge hyperparameter $\beta=0.5$; setting $\zeta=0.1$ gives $\epsilon_g=0.05$ as the optimal expert with the performance similar to BayesAoA-ES for $N=8$ and $\sigma^2=10^{-6}$. The proposed HedgeBayesAoA-ES gives us a way to tune the thresholds according to the number of receivers $N$ and the noise variance $\sigma^2$ dynamically unlike a BayesAoA-ES method where the algorithm should be deployed with a single $\epsilon_g$ that may not be appropriate to get the best accuracy and the least computation for all $N$. The deep learning (DL) based AoA techniques also need to retrain the models with the change in the set up like noise variance and the number of receivers used at the receiver, thus cannot be used in an online fashion. Once deployed, the already trained DL model does not perform well in a different setup. The proposed HedgeBayesAoA-ES helps to overcome this issue.

    \section{Conclusion}
In summary, we have proposed an SMBO based Bayesian approach for AoA estimation technique where the current sampling depends on the outcome of the previous samplings. The proposed method achieves an accuracy of $92\%$ with $80.7\%$ saving in computation compared to the brute-force method. The method is insensitive towards initialization, unlike EM or SAGE. We further propose to use a Hedge type solution to pick the best $\epsilon_g$ for a given channel condition and a receiver configuration. In this dynamic environment, our method is suitable for AoA estimation in an online manner. Further, the proposed method has less complexity and needs lesser computing power than traditional deep learning AoA estimation techniques.
 	\bibliographystyle{IEEEtran}
    \bibliography{library.bib}

\begin{thebibliography}{10}
\providecommand{\url}[1]{#1}
\csname url@samestyle\endcsname
\providecommand{\newblock}{\relax}
\providecommand{\bibinfo}[2]{#2}
\providecommand{\BIBentrySTDinterwordspacing}{\spaceskip=0pt\relax}
\providecommand{\BIBentryALTinterwordstretchfactor}{4}
\providecommand{\BIBentryALTinterwordspacing}{\spaceskip=\fontdimen2\font plus
\BIBentryALTinterwordstretchfactor\fontdimen3\font minus
  \fontdimen4\font\relax}
\providecommand{\BIBforeignlanguage}[2]{{%
\expandafter\ifx\csname l@#1\endcsname\relax
\typeout{** WARNING: IEEEtran.bst: No hyphenation pattern has been}%
\typeout{** loaded for the language `#1'. Using the pattern for}%
\typeout{** the default language instead.}%
\else
\language=\csname l@#1\endcsname
\fi
#2}}
\providecommand{\BIBdecl}{\relax}
\BIBdecl

\bibitem{van2004optimum}
H.~L. Van~Trees, \emph{Optimum array processing: Part IV of detection,
  estimation, and modulation theory}.\hskip 1em plus 0.5em minus 0.4em\relax
  John Wiley \& Sons, 2004.

\bibitem{capon1969high}
J.~Capon, ``High-resolution frequency-wavenumber spectrum analysis,''
  \emph{Proceedings of the IEEE}, vol.~57, no.~8, pp. 1408--1418, 1969.

\bibitem{schmidt1986multiple}
R.~Schmidt, ``Multiple emitter location and signal parameter estimation,''
  \emph{IEEE transactions on antennas and propagation}, vol.~34, no.~3, pp.
  276--280, 1986.

\bibitem{barabell1983improving}
A.~Barabell, ``Improving the resolution performance of eigenstructure-based
  direction-finding algorithms,'' in \emph{ICASSP'83. IEEE International
  Conference on Acoustics, Speech, and Signal Processing}, vol.~8.\hskip 1em
  plus 0.5em minus 0.4em\relax IEEE, 1983, pp. 336--339.

\bibitem{roy1989esprit}
R.~Roy and T.~Kailath, ``Esprit-estimation of signal parameters via rotational
  invariance techniques,'' \emph{IEEE Transactions on acoustics, speech, and
  signal processing}, vol.~37, no.~7, pp. 984--995, 1989.

\bibitem{shan1985spatial}
T.-J. Shan, M.~Wax, and T.~Kailath, ``On spatial smoothing for
  direction-of-arrival estimation of coherent signals,'' \emph{IEEE
  Transactions on Acoustics, Speech, and Signal Processing}, vol.~33, no.~4,
  pp. 806--811, 1985.

\bibitem{ziskind1988maximum}
I.~Ziskind and M.~Wax, ``Maximum likelihood localization of multiple sources by
  alternating projection,'' \emph{IEEE Transactions on Acoustics, Speech, and
  Signal Processing}, vol.~36, no.~10, pp. 1553--1560, 1988.

\bibitem{chung2002doa}
P.~J. Chung and J.~F. B{\"o}hme, ``Doa estimation using fast em and sage
  algorithms,'' \emph{Signal Processing}, vol.~82, no.~11, pp. 1753--1762,
  2002.

\bibitem{malioutov2005sparse}
D.~Malioutov, M.~Cetin, and A.~S. Willsky, ``A sparse signal reconstruction
  perspective for source localization with sensor arrays,'' \emph{IEEE
  transactions on signal processing}, vol.~53, no.~8, pp. 3010--3022, 2005.

\bibitem{bnilam2020angle}
N.~Bnilam, E.~Tanghe, J.~Steckel, W.~Joseph, and M.~Weyn, ``Angle: Angular
  location estimation algorithms,'' \emph{IEEE Access}, vol.~8, pp.
  14\,620--14\,629, 2020.

\bibitem{bnilam2020loray}
N.~BniLam, D.~Joosens, M.~Aernouts, J.~Steckel, and M.~Weyn, ``Loray: Aoa
  estimation system for long range communication networks,'' \emph{IEEE
  Transactions on Wireless Communications}, 2020.

\bibitem{huang2018deep}
H.~Huang, J.~Yang, H.~Huang, Y.~Song, and G.~Gui, ``Deep learning for
  super-resolution channel estimation and doa estimation based massive mimo
  system,'' \emph{IEEE Transactions on Vehicular Technology}, vol.~67, no.~9,
  pp. 8549--8560, 2018.

\bibitem{Haykin1991AdvancesIS}
S.~Haykin, ``Advances in spectrum analysis and array processing,'' 1991.

\bibitem{fessler1994space}
J.~A. Fessler and A.~O. Hero, ``Space-alternating generalized
  expectation-maximization algorithm,'' \emph{IEEE Transactions on signal
  processing}, vol.~42, no.~10, pp. 2664--2677, 1994.

\bibitem{bergstra2012random}
J.~Bergstra and Y.~Bengio, ``Random search for hyper-parameter optimization.''
  \emph{Journal of machine learning research}, vol.~13, no.~2, 2012.

\bibitem{kennedy1995particle}
J.~Kennedy and R.~Eberhart, ``Particle swarm optimization,'' in
  \emph{Proceedings of ICNN'95-international conference on neural networks},
  vol.~4.\hskip 1em plus 0.5em minus 0.4em\relax IEEE, 1995, pp. 1942--1948.

\bibitem{bergstra2011algorithms}
J.~Bergstra, R.~Bardenet, Y.~Bengio, and B.~K{\'e}gl, ``Algorithms for
  hyper-parameter optimization,'' in \emph{25th annual conference on neural
  information processing systems (NIPS 2011)}, vol.~24.\hskip 1em plus 0.5em
  minus 0.4em\relax Neural Information Processing Systems Foundation, 2011.

\bibitem{freund1997decision}
Y.~Freund and R.~E. Schapire, ``A decision-theoretic generalization of on-line
  learning and an application to boosting,'' \emph{Journal of computer and
  system sciences}, vol.~55, no.~1, pp. 119--139, 1997.

\end{thebibliography}
\end{document}